\documentstyle[aps,preprint]{revtex}
\begin{document}
\author{Remo Garattini}
\address{Facolt\`a di Ingegneria, Universit\`a di Bergamo\\
Viale Marconi, 5, 24044 Dalmine (Bergamo) Italy\\
e-mail: Garattini@mi.infn.it}
\title{Vacuum Energy Estimates in Quantum Gravity and the Wheeler-DeWitt Equation}
\date{April 1, 1996}
\maketitle

\begin{abstract}
In the Wheeler-DeWitt framework, by a gauge fixing procedure, we set up a
scheme to recover a Schr\"odinger type equation, living in the orbits space
with the {\it lapse} function as evolution parameter. By means of the
associated stationary equation, we have the possibility of calculating
quantum corrections to classical quantities. The Schwarzschild wormhole case
is discussed as an example of application.
\end{abstract}

\section{Introduction}

\label{p0}

Any attempt to apply the fundamentals of Quantum Mechanics to a complicated
theory such as gravity is of central importance; as regards this problem it
is enough thinking of any kind of physical situation whatsoever to convince
ourselves that the Quantum Principles constitute an intrinsic property of
nature. However a satisfactory model of Quantum Gravity is still far from
realization. Nevertheless, the basic principles that lead to a quantization
scheme for theories such as, e.g., quantum electrodynamics, are very
encouraging; in particular the Feynman sum over histories is a powerful
method to deal with quantities like two point correlation functions,
propagators, etc $\ldots $ . Although the sum over histories maintains a
covariant form, we encounter two problems with this approach (we omit
renormalization and the unboundedness from below problems):

\begin{description}
\item[a)]  Due to the integration over $4D$, one looses any meaning for time.
\end{description}

\begin{description}
\item[b)]  If we decide to develop our formalism in the Euclidean time, some
topologies suggest that the manifold has to be compact or compactified;
usually this means that every result has to be understood from a
thermodynamical point of view, i.e., at equilibrium: when time ceases to
have a meaning.
\end{description}

In both cases, the result is the disappearance of the dynamics. No time
evolution can be recognized or extracted. On the other side, the canonical
procedure, via the hamiltonian formulation, introduces a natural splitting
between space and time, described by the {\it lapse} and {\it shift}
functions, denoted as 
\begin{equation}
\label{ii0}\left( N,N^i\right) =N^\mu . 
\end{equation}
respectively. In terms of these functions, the line element becomes 
\begin{equation}
\label{ii1}ds_{}^2=g_{\mu \nu }\left( x\right) dx^\mu dx^\nu =\left(
-N^2+N_iN^i\right) dt^2+2N_jdtdx^j+g_{ij}dx^idx^j. 
\end{equation}
Unfortunately, when we make this choice, every dynamical quantity is
constrained to zero, both at the classical and at the quantum level. The
quantum constraint for the {\it lapse} function is known as Wheeler-DeWitt
equation (WDW) and is represented by the equation: 
\begin{equation}
\label{ii2}{\cal H}\Psi =0. 
\end{equation}
Once again, any time information is entangled inside the formalism.
Accordingly, any possible energy definition is meaningless within this
framework. Since a basic feature of a Quantum Theory is the existence of a
Ground State and this is related to our capability of defining and
calculating{\it \ energy}, one sees the necessity of a Schr\"odinger
equation. Actually, within the WDW framework, it is possible to recover an
internal time by a temporary suspension of the constraint and it is well
known from Quantum Cosmology that this procedure gives rise to a
Schr\"odinger equation. In practice, to see how things work, we introduce a
gauge fixing that foliates spacetime. But it is well known that the
consequence of a gauge fixing is the appearance of the Faddeev-Popov
determinant and of an integration over the volume group. The volume group
integration represents the sum over all possible lapses and this restores
the gauge invariance. Anyway, this is not enough, because we have a very
large variety of configurations available from the 3-space, and for this
reason, we focus our attention on wormhole physics, in the sense of ref.\cite
{MTW}. To examine how wormholes affect the ground state energy
configuration, we suggest that matter be excluded from this picture:
although this representation looks unreal, because matter is intimately
related to gravity, it seems to be important understanding the effect of
pure gravity on some primordial configurations of the Universe. To this aim,
it has been suggested that Wheeler's spacetime foam could be a good
candidate for the gravitational ground state \cite{Hawking}. It is clear,
now, how wormholes can enter into the vacuum state: spacetime foam can be
approximated by a multi-wormholes configuration. However, in this paper we
begin with the analysis of a single wormhole to understand what kind of
consequences on the energy defined on every space-time slice we can observe.
But, how can we probe vacuum ? Despite the covariant philosophy, we use the
Hamiltonian in its unconstrained version. The plan of the paper is the
following: in section \ref{p1}, we introduce the path integral
representation; in section \ref{p2}, we propose the wave functional needed
to study the possible ground state; in section \ref{p3}, we show the
properties of the variational calculation, in section \ref{p4}, we summarize
and conclude.

\section{Path Integral Representation}

\label{p1}

If we take a glance to $\left( \ref{ii2}\right) $, we see that the path
integral expression 
\begin{equation}
\label{i0}
\begin{array}{c}
\int 
{\cal D}g_{\mu \nu }\exp \left( iS_g\left[ g_{\mu \nu }\right] \right)
\qquad \text{(Lorentzian)} \\  \\ 
\int {\cal D}g_{\mu \nu }\exp \left( -I_g\left[ g_{\mu \nu }\right] \right)
\qquad \text{(Euclidean)} 
\end{array}
\end{equation}
generates solutions to the WDW equation and the momentum constraints 
\begin{equation}
\label{i1}{\cal H}\Psi =0\qquad {\cal H}_i\Psi =0. 
\end{equation}
In $\left( \ref{i0}\right) $, $S_g\left[ g_{\mu \nu }\right] $ $\left(
I_g\left[ g_{\mu \nu }\right] \right) $ is the Lorentzian (Euclidean)
Einstein-Hilbert action for gravity that is stationary under variations of
the metric vanishing on the boundary 
\begin{equation}
\label{i2}
\begin{array}{c}
S_g\left[ g_{\mu \nu }\right] = 
\frac 1{16\pi G}\int\limits_{{\cal M}}d^4x\sqrt{-g}R+\frac 1{8\pi G}%
\int\limits_{{\cal \partial M}}d^3x\sqrt{h}K\qquad \text{(Lorentzian)} \\  
\\ 
I_g\left[ g_{\mu \nu }\right] =-\frac 1{16\pi G}\int\limits_{{\cal M}}d^4x%
\sqrt{g}R-\frac 1{8\pi G}\int\limits_{{\cal \partial M}}d^3x\sqrt{h}K\qquad 
\text{(Euclidean).} 
\end{array}
\end{equation}
Consider, now, expression $\left( \ref{i0}\right) $, in its lorentzian form:
the wave function constructed from the path integral is represented,
explicitly, by \cite{Halliwell}: 
\begin{equation}
\label{aa1}\Psi \left[ \widetilde{g}_{ij}\right] =\int_\gamma {\cal D}N\int 
{\cal D}g_{ij}{\cal D}\pi ^{ij}\triangle \left[ g_{ij},\pi ^{ij},N\right]
\delta \left( \dot N-\chi \left( g_{ij},\pi ^{ij},N\right) \right) \exp
\left( iS\left[ g_{ij},\pi ^{ij},N\right] \right) , 
\end{equation}
where $S$ is the action, $N$ is the lapse function and $\chi $ is an
arbitrary function entering the gauge-fixing condition in the argument of
the delta function. $\triangle $ is the associated Faddeev-Popov
determinant. Equation $\left( \ref{aa1}\right) $ defines an invariant class
of paths to integrate over. Suppose to consider the gauge $\chi =0=\dot N,$
then $\triangle $ may be shown to equal a constant and after integrating
over the momenta $\pi ^{ij}$, the wave function becomes 
\begin{equation}
\label{aa1b}\Psi \left[ \widetilde{g}_{ij}\right] =\int_\gamma {\cal D}N\int 
{\cal D}g_{ij}\exp \left( iS\left[ g_{ij},N\right] \right) ; 
\end{equation}
the integral over $g$ defines a wave function $\Psi \left[ \widetilde{g}%
_{ij},N\right] $ which satisfies the familiar Schr\"odinger equation 
\begin{equation}
\label{aa2}{\cal H}\Psi =i\frac{\partial \Psi }{\partial N}. 
\end{equation}
For completeness, we report eq.$\left( \ref{aa2}\right) $ in the Euclidean
signature 
\begin{equation}
\label{aa2a}{\cal H}\Psi =-\frac{\partial \Psi }{\partial N} 
\end{equation}
The remaining integral is over the constant value of $N$ 
\begin{equation}
\label{aa3}\Psi \left[ \widetilde{g}_{ij}\right] =\int_\gamma dN\Psi \left[
g_{ij},N\right] . 
\end{equation}
$\Psi \left[ \widetilde{g}_{ij}\right] ,$ defined by $\left( \ref{aa3}%
\right) $, will satisfy the WDW equation, if the range is chosen to be from $%
-\infty $ to $+\infty $, or if the $N$ contour is closed. The action takes
the form 
\begin{equation}
\label{aa4}S\left[ g_{ij},N\right] =\frac 1{16\pi G}\int_0^1dtN\int d^3x%
\sqrt{g}\left[ K_{ij}K^{ij}-K^2-\!\ ^3R\right] , 
\end{equation}
where $K_{ij}=\frac 1{2N}\left( N_{i|j}+N_{j|i}-g_{ij,0}\right) $ is called
the second fundamental form and $K=g^{ij}K_{ij}$ is the trace associated to $%
K_{ij}.$ The range of the $t$ integration may be taken from $0$ to $1$, by
shifting t and by scaling the lapse function. Consider the WDW operator
introduced in $\left( \ref{ii2}\right) $ or in its unconstrained form $%
\left( \ref{aa2}\right) $, whose expression is 
\begin{equation}
\label{aa5}{\cal H}=G_{ijkl}\frac{\pi _{}^{ij}\pi ^{kl}}{\sqrt{g}}-\sqrt{g}%
R, 
\end{equation}
where 
$$
G_{ijkl}=g_{ik}g_{jl}+g_{il}g_{jk}-2g_{ij}g_{kl} 
$$
is called the inverse supermetric. Since the scalar curvature behaves as a
potential and since we are interested to the Ground State problem, we
perform an expansion of the 3-metric around a fixed background: in
particular, let us choose to work with stationary spherical symmetric
metrics. If we consider the expansion up to the second order, $\left( \ref
{aa5}\right) $ becomes a harmonic oscillator. Therefore, after some
algebraic manipulations we obtain 
\begin{equation}
\label{aa6}{\cal H}=G_{ijkl}\frac{\pi _{}^{ij}\pi ^{kl}}{\sqrt{g}}-\sqrt{g}%
G_{}^{ijkl}h_{i\alpha }O_{jk}^{\alpha \beta }h_{\beta l}, 
\end{equation}
where $O_{jk}^{\alpha \beta }$ is an operator containing second order
derivatives and%
$$
G_{}^{ijkl}=g^{ik}g^{jl}+g^{il}g^{jk}-2g^{ij}g^{kl} 
$$
is called the supermetric (an explicit expression and derivation of $\left( 
\ref{aa6}\right) $ being discussed in \cite{Remo}). This implies that the
wave function, satisfying the approximated problem, has to be changed too.
By defining\footnote{%
Since we are working with static configurations, all contributions coming
from the temporal part, i.e. from the {\it Kinetic} term, have to be
integrated out from the space, but not from the path integral.} 
\begin{equation}
\label{aa7}\omega =\frac 1{16\pi G}\int d^3x\sqrt{g}\left[
K_{ij}K^{ij}-K^2\right] =\frac 1{16\pi G}\int d^3x\sqrt{g}\left[
G^{ijkl}K_{ij}K_{kl}\right] , 
\end{equation}
and taking 
\begin{equation}
\label{aa8}K_{ij}=-\frac 1{2N}\dot g_{ij}\simeq -\frac 1{2N}\dot h_{ij}, 
\end{equation}
we obtain 
\begin{equation}
\label{aa9}\omega =\frac 1{16\pi G}\int d^3x\sqrt{g}\left[ G^{ijkl}\frac 1{%
4N^2}\dot h_{ij}\dot h_{kl}\right] . 
\end{equation}
In $\left( \ref{aa8}\right) ,$ the compatibility with the choice of the
background metric is evident. We recall, in fact, that the spherical
symmetric metrics (s.s.m.) have their (non-zero) elements on the principal
diagonal, and it seems natural that along non principal directions something
will fluctuate. This choice is completely in agreement with the homogeneous
minisuperspace framework. As to eq.$\left( \ref{aa9}\right) ,$ we can see
quantum fluctuations entering it, and this means that $\omega $ is a
functional of $h$; actually, when we consider the operator appearing in $%
\left( \ref{aa6}\right) $, it is immediate to recognize that, if we solve
the following eigenvalue problem 
\begin{equation}
\label{aa10}O_j^\alpha h_\alpha ^i=\lambda h_j^i, 
\end{equation}
a basis of eigenfunctions can be used to express the fields: so that the
integration in $\left( \ref{aa9}\right) $ is justified.\footnote{%
The operator appearing in $\left( \ref{aa6}\right) $ is formally the same
that we find when a one loop expansion is performed around a certain fixed
background in $4D$. Nevertheless, in this approach, the associated equation $%
\left( \ref{aa10}\right) $ is defined in the three dimensional subspace of
the initial covariant space.}

On the previous grounds, the wave function 
\begin{equation}
\label{aa11}\int {\cal D}g_{ij}\exp i\left( \int_0^1dtN\omega -\frac 1{16\pi
G}\int_0^1dtN\int d^3x\sqrt{g}\!\ \left( ^3R\right) \right) 
\end{equation}
becomes 
\begin{equation}
\label{aa12a}\exp iN\omega \int {\cal D}g_{ij}\exp -i\left( \frac 1{16\pi G}%
N\int d^3x\sqrt{g}\!\ \left( ^3R\right) \right) =\exp iN\omega \Phi \left(
N,g_{ij}\right) . 
\end{equation}
However, by considering rescaled fields with respect to the lapse function,
we obtain 
\begin{equation}
\label{aa12}\exp iN\omega \Phi \left( N,g_{ij}\right) =\exp iN\omega \Phi
\left( g_{ij}^{\prime }\right) , 
\end{equation}
where $g_{ij}^{\prime }=g_{ij}^{}/\sqrt{N}$ .

\begin{description}
\item[Remark ]  - Actually, equation $\left( \ref{aa6}\right) $ together
with eq.$\left( \ref{aa10}\right) $ contain a substructure that comes up
when a convenient decomposition is made with the help of ultralocal metric.
Indeed, ultralocal metric is a more general kind of supermetric and within
such a framework we can distinguish three types of operators describing spin
2, spin 1 and spin 0 fields; each of these components is orthogonal to the
others in the sense of ultralocality \cite{Remo}.
\end{description}

\section{The Schr\"odinger Equation: another point of view}

\label{p2}

Instead of attacking the problem of the solutions of the WDW equation and
consequently of the associated Schr\"odinger equation, we observe that 
\begin{equation}
\label{a1}{\cal H}\Psi \left[ \widetilde{g}_{ij}\right] =0=\int_\gamma
dN\left( i\frac{\partial \Psi \left[ g_{ij},N\right] }{\partial N}\right) . 
\end{equation}
is not forbidden by $\left( \ref{ii2}\right) ,$ provided the integration
contour is invariant under reparametrization: and this is true for closed
contours or infinite end points. While the r.h.s. of $\left( \ref{a1}\right) 
$ vanishes because of boundary conditions, the l.h.s. of the same equation
needs a specification: if we have a look at $\left( \ref{a1}\right) $, we
immediately see that, by defining 
\begin{equation}
\label{a2}\Psi \left[ \widetilde{g}_{ij}\right] :=\int_\gamma dN\Psi \left[
g_{ij},N\right] , 
\end{equation}
we have formally 
\begin{equation}
\label{a3}{\cal H}\Psi \left[ \widetilde{g}_{ij},N\right] =i\frac{\partial
\Psi \left[ g_{ij},N\right] }{\partial N}, 
\end{equation}
which is a Schr\"odinger equation; and the invariance under time
reparametrization survives if and only if the sum over all lapses of $\Psi
\left[ \widetilde{g}_{ij},N\right] $ recovers the WDW equation. Suppose,
now, to separate the variables 
\begin{equation}
\label{a3a}\Psi \left[ \widetilde{g}_{ij},N\right] =F\left( N\right) \Phi
\left( g_{ij}\right) ; 
\end{equation}
then $\left( \ref{a3}\right) $ splits into: 
\begin{equation}
\label{a4}
\begin{array}{c}
{\cal H}\Phi \left[ g_{ij}\right] =\omega \Phi \left[ g_{ij}\right] \\  \\ 
i\frac{dF\left[ N\right] }{dN}=\omega F\left[ N\right] 
\end{array}
\end{equation}
and $\left( \ref{a3a}\right) $ becomes%
$$
\Psi \left[ \widetilde{g}_{ij},N\right] =\exp -iN\omega \Phi \left(
g_{ij}\right) , 
$$
which is eq.$\left( \ref{aa12}\right) $, up to a phase factor. Taking
advantage of this temporary separation between lapse and three-fields, one
immediately sees a procedure for calculating the spectrum of ${\cal H}$ at
fixed $N.$ That problem was studied in \cite{Garay}. Our present purpose is
understanding the vacuum structure, and in section \ref{p1} an assumption
about the ground state was made by us. In other words, instead of solving
directly the stationary equation, we adopt a variational procedure with
gaussian wave functional as trial functional.

\section{The Gaussian Wave Functionals: an application to the Schwarzschild
wormhole}

\label{p3}

Let us recall the basic rules on gaussian wave functionals.

The action of the operator $h_{ij}^{}$on $|\Phi \rangle =\Phi \left[
h_{ij}^{}\left( \overrightarrow{x}\right) \right] $ is realized by 
\begin{equation}
\label{c1}h_{ij}^{}\left( x\right) |\Phi \rangle =h_{ij}^{}\left( x\right)
\Phi \left[ h_{ij}^{}\left( \overrightarrow{x}\right) \right] . 
\end{equation}
The action of the operator $\pi _{ij}^{}$on $|\Phi \rangle $, in general, is

\begin{equation}
\label{c2}\pi _{ij}^{}\left( x\right) |\Phi \rangle =-i\frac \delta {\delta
h_{ij^{}}^{}\left( x\right) }\Phi \left[ h_{ij}^{}\left( \overrightarrow{x}%
\right) \right] . 
\end{equation}
The inner product is defined by the functional integration: 
\begin{equation}
\label{c3}\left\langle \Phi _1\mid \Phi _2\right\rangle =\int \left[ {\cal D}%
h_{ij}^{}\left( x\right) \right] \Phi _1^{*}\left\{ h_{ij}^{}\right\} \Phi
_2\left\{ h_{kl}^{}\right\} , 
\end{equation}
and the energy eigenstates satisfy the Schr\"odinger equation: 
\begin{equation}
\label{c4}\int d_{}^3x{\cal H}\left\{ -i\frac \delta {\delta
h_{ij^{}}^{}\left( x\right) },h_{ij}\left( x\right) \right\} \Phi \left\{
h_{ij}^{}\left( \overrightarrow{x}\right) \right\} =\omega \Phi \left\{
h_{ij}^{}\left( \overrightarrow{x}\right) \right\} , 
\end{equation}
where ${\cal H}\left\{ -i\frac \delta {\delta h_{ij^{}}^{}\left( x\right) }%
,h_{ij}\left( x\right) \right\} $ is the Hamiltonian density. Instead of
solving $\left( \ref{c4}\right) $, which is of course impossible, we can
formulate the same problem by means of a variational principle. We demand
that 
\begin{equation}
\label{c6}\frac{\left\langle \Phi _{}^{}\mid H\mid \Phi _{}\right\rangle }{%
\left\langle \Phi _{}^{}\mid \Phi _{}\right\rangle }=\frac{\int \left[ {\cal %
D}g_{ij}^{}\left( x\right) \right] \int d_{}^3x\Phi _1^{*}\left\{
g_{ij}^{}\right\} {\cal H}\Phi \left\{ g_{kl}^{}\right\} }{\int \left[ {\cal %
D}g_{ij}^{}\left( x\right) \right] \mid \Phi \left\{ g_{ij}^{}\right\} \mid
_{}^2} 
\end{equation}
be stationary against arbitrary variations of $\Phi \left\{ h_{ij}^{}\left( 
\overrightarrow{x}\right) \right\} $. The form of $\left\langle \Phi
_{}^{}\mid H\mid \Phi _{}\right\rangle $ can be computed as follows. We
define normalized mean values 
\begin{equation}
\label{c7}\bar g_{ij}^{}\left( x\right) =\frac{\int \left[ {\cal D}%
g_{ij}^{}\left( x\right) \right] \int d_{}^3xg_{ij}^{}\left( x\right) \mid
\Phi \left\{ g_{ij}^{}\left( x\right) \right\} \mid _{}^2}{\int \left[ {\cal %
D}g_{ij}^{}\left( x\right) \right] \mid \Phi \left\{ g_{ij}^{}\right\} \mid
_{}^2}, 
\end{equation}
\begin{equation}
\label{c8}\bar g_{ij}^{}\left( x\right) \text{ }\bar g_{kl}^{}\left(
x\right) +K_{ijkl^{}}^{}\left( \overrightarrow{x},\overrightarrow{y}\right) =%
\frac{\int \left[ {\cal D}g_{ij}^{}\left( x\right) \right] \int
d_{}^3xg_{ij}^{}\left( x\right) g_{kl}^{}\left( y\right) \mid \Phi \left\{
g_{ij}^{}\left( x\right) \right\} \mid _{}^2}{\int \left[ {\cal D}%
g_{ij}^{}\left( x\right) \right] \mid \Phi \left\{ g_{ij}^{}\right\} \mid
_{}^2}. 
\end{equation}
It follows that%
$$
\int \left[ {\cal D}h_{ij}^{}\left( x\right) \right] \left( g_{ij}^{}\left(
x\right) -\bar g_{ij}^{}\left( x\right) \right) \mid \Phi \left\{
g_{ij}^{}\left( x\right) \right\} \mid _{}^2=0 
$$
by translation invariance of the measure%
$$
\int \left[ {\cal D}h_{ij}^{}\left( x\right) \right] h_{ij}^{}\left(
x\right) \mid \Phi \left\{ g_{ij}^{}\left( x\right) +\bar g_{ij}^{}\left(
x\right) \right\} \mid _{}^2=0 
$$
\begin{equation}
\label{c9}\Longrightarrow \int \left[ {\cal D}h_{ij}^{}\left( x\right)
\right] h_{ij}^{}\left( x\right) \mid \Phi \left\{ h_{ij}^{}\left( x\right)
\right\} \mid _{}^2=0 
\end{equation}
and $\left( \ref{c8}\right) $ becomes 
\begin{equation}
\label{c10}
\begin{array}{c}
\int \left[ 
{\cal D}h_{ij}^{}\left( x\right) \right] \int d_{}^3xh_{ij}^{}\left(
x\right) h_{kl}^{}\left( y\right) \mid \Phi \left\{ h_{ij}^{}\left( x\right)
\right\} \mid _{}^2= \\  \\ 
K_{ijkl^{}}^{}\left( \overrightarrow{x},\overrightarrow{y}\right) \int
\left[ {\cal D}h_{ij}^{}\left( x\right) \right] \mid \Phi \left\{
h_{ij}^{}\right\} \mid _{}^2 
\end{array}
\end{equation}
Rather than applying the variational principle arbitrarily, the gaussian 
{\it Ansatz} is used, according to which in the beginning of this calculus
one has to replace the previous general formulas by 
\begin{equation}
\label{c11}\Phi _\alpha \left[ h_{ij}^{}\left( \overrightarrow{x}\right)
\right] ={\cal N}\exp \left\{ -\frac N{4l_p^2}\left\langle \left( g-%
\overline{g}\right) K_{}^{-1}\left( g-\overline{g}\right) \right\rangle
_{x,y}^{\bot }+\ldots \ldots \right\} . 
\end{equation}
$\left\langle .,.\right\rangle _{x,y}^{\bot }$ means that we are integrating
over $x$ and $y$ in the volume and considering only the physical fields are
considered, that is, traceless and divergenceless fields (spin 2). Actually,
in $\left( \ref{c11}\right) $ we should have to consider the rescaled fields
with respect to the lapse function, like in $\left( \ref{aa12}\right) $.
With this choice and with formulae $\left( \ref{c9},\ref{c10}\right) $, the
one loop-like Hamiltonian can be written as 
\begin{equation}
\label{c12}\omega _{}^{\bot }=\frac 1{4l_p^2}\int_{{\cal M}}^{}d_{}^3x\sqrt{g%
}G_\alpha ^{ijkl}\left[ K_{}^{-1\bot }\left( x,x\right) _{ijkl}+\left(
\triangle _2^{}\right) _j^aK_{}^{\bot }\left( x,x\right) _{iakl}\right] 
\end{equation}
where the first term in square brackets comes from the kinetic part, while
the second comes from the expansion of the $^3R$ up to second order; $%
G_\alpha ^{ijkl}$ is called the ultralocal metric which describes the WDW
metric with the choice $\alpha =1$ and $l_p$ is the Planck length. The Green
function $K_{}^{\bot }\left( x,x\right) _{iakl}$ can be represented as 
\begin{equation}
\label{c13}K_{}^{\bot }\left( x,x\right) _{iakl}:=\sum_N\frac{h_{ia}^{\bot
}\left( x\right) h_{kl}^{\bot }\left( y\right) }{2\lambda _N\left( p\right) }%
, 
\end{equation}
where $h_{ia}^{\bot }\left( x\right) $ are the eigenfunctions relative to $%
\left( \ref{aa10}\right) ,$ and $\lambda _N\left( p\right) $ are infinite
variational parameters. In formula $\left( \ref{c12}\right) $ we have
written the Spin 2 contribution to the energy density alone; expressions
like $\left( \ref{c12}\right) $ exist for the Spin 1 and Spin 0 terms of $%
{\cal H}$. As an application we consider the energy evaluations coming from
the Schwarzschild and Flat background. If 
\begin{equation}
\label{c14}
\begin{array}{c}
g_{ij}=g_{ij}^{\left( S\right) }+h_{ij} \\  
\\ 
g_{ij}=g_{ij}^{\left( F\right) }+h_{ij}, 
\end{array}
\end{equation}
then the stationary Schr\"odinger equations are respectively 
\begin{equation}
\label{c15}{\cal H}_S\Phi \left( h_{ij}\right) =\omega _S\Phi \left(
h_{ij}\right) 
\end{equation}
for the curved background, and 
\begin{equation}
\label{c16}{\cal H}_F\Phi \left( h_{ij}\right) =\omega _F\Phi \left(
h_{ij}\right) 
\end{equation}
for the flat one. Due to the ultraviolet divergences, it is better to
calculate the difference between $\left( \ref{c15}\right) $ and $\left( \ref
{c16}\right) :$%
\begin{equation}
\label{c17}\left( {\cal H}_S-{\cal H}_F\right) \Phi \left( h_{ij}\right)
=\left( \omega _S-\omega _F\right) \Phi \left( h_{ij}\right) . 
\end{equation}
Since we are performing a difference between two different manifolds and
this operation is ill defined, the same evaluation can be done by
approaching flat space in the limit of $M$ tending to zero. Then the result,
up to quadratic order, is: 
\begin{equation}
\label{c18}\omega _S-\omega _F=\triangle \omega \left( M\right) =-\frac V{%
2\pi ^2}\left( \frac{3MG}{r_0^3}\right) ^2\frac 1{16}\ln \left( \frac{%
\Lambda ^2}{\frac{3MG}{r_0^3}}\right) , 
\end{equation}
where $M$ is the Schwarzschild mass parameter, $V$ is the volume of the
system, $\Lambda $ is the UV cut-off and $r_0$ represents the closest radius
reachable compatible with quantum effects. In terms of the wormhole volume
eq.$\left( \ref{c18}\right) $ becomes 
\begin{equation}
\label{c19}\triangle \omega \left( M\right) =-\frac V{2\pi ^2}\left( \frac{%
4\pi MG}{V_S}\right) ^2\frac 1{16}\ln \left( \frac{V_S\Lambda ^2}{4\pi MG}%
\right) . 
\end{equation}

\section{Conclusion, Open problems and Outlooks}

\label{p4}

Although the framework presented in section $\ref{p2}$ and $\ref{p3}$ need
more rigorous arguments, equations $\left( \ref{a4}\right) $ and $\left( \ref
{aa12}\right) $ define a calculation scheme useful for the energy estimates.
Indeed, what we have built is a calculation framework useful to estimate the
contribution of the three field quantum fluctuation around a classical
solution of the Einstein equations, even though, when we turn to the
Hamiltonian, the {\it lapse} is more general. It is important to point out
that in previous formulae no mention has been made to the possibility of
having contributions from the asymptotic energy or ${\cal ADM}$ mass, which
in the particular case of the Schwarzschild background is identified with
the parameter $M$ entering the line element expression. Since $E_{{\cal ADM}%
}^{}$ corresponds to the total mass, which is conjugate to the time
separation at spatial infinity, a problem of matching with the {\it time}
appears into eq.$\left( \ref{aa2}\right) $ . Indeed, by defining an
appropriate asymptotic wavefunctional $\Psi =\Psi \left[ T,^3{\cal G}\right]
,$ which obeys the Schr\"odinger equation \cite{Carlip} 
\begin{equation}
\label{co1}M\Psi =i\frac{\partial \Psi }{\partial T}, 
\end{equation}
we can observe a difference with respect to the wave functional used
throughout the paper, which obeys eq.$\left( \ref{aa2}\right) .$ According
to ref.\cite{Carlip}, a possibility is 
\begin{equation}
\label{co2}N\rightarrow T\qquad \text{when\qquad }r\rightarrow \infty . 
\end{equation}
Then the complete wavefunction assumes the form 
\begin{equation}
\label{co3}\Psi \left[ \widetilde{g}_{ij}\right] :=\int_\gamma dN\Psi \left[
g_{ij},N\right] =\int_\gamma dN\ e^{-i\left( N\omega +TM\right) }\Phi \left(
g_{ij}\right) 
\end{equation}
and $\left( \ref{c19}\right) $ becomes 
\begin{equation}
\label{co4}M-\frac V{2\pi ^2}\left( \frac{4\pi MG}{V_S}\right) ^2\frac 1{16}%
\ln \left( \frac{V_S\Lambda ^2}{4\pi MG}\right) . 
\end{equation}

Throughout this paper, apart a reparametrization invariance condition, no
mention has been made about the lapse boundary conditions. The reason is
that a part of this scheme is independent from the boundary choices: in
fact, the average calculation performed with the help of the variational
approach involves only the spatial part, and only at the end of this
operation one recovers invariance by integrating over all possible lapses.
At this point a conjecture on how thermal problems can be described by pure
dynamics can be made. In particular, we could ask ourselves how the
constraint equation (WDW) affects the Hawking temperature and in which way
one can extract the characteristic relation $T_{eq}=\frac 1{8\pi M}$ . Let
us consider this calculation scheme: because the eigenvalue obtained by the
variational procedure is independent of the lapse, we could, in principle,
compare the results $\left( \ref{c19}\right) $ of the Lorentzian and
Euclidean formulations; but it is well known that the introduction of an
Euclidean time involves a temperature in the Schwarzschild sector and
usually this is identified with the Hawking temperature to avoid the conical
singularity. But in the Lorentzian signature we do not need introducing
neither a temperature nor an imaginary time, which means that we can
understand what is happening in this particular thermalization process.
Another open issue is the appearing of a discrete spectrum in the
Schwarzschild as well as in the Schwarzschild-DeSitter sector \cite{GPY},
together which the continuum one. Since the discrete spectrum was obtained
by means of an Euclidean path integral, the usual interpretation of the
associated eigenvalues is the possibility of having instability from hot
flat space. From the point of view of this paper, it could be interesting to
verify the existence of a discrete spectrum in this situation and to perform
a comparison between the Euclidean and Lorentzian sectors \cite{Remo1}.

\section{Acknowledgments}

I wish to thank M. Cavagli\`a, H.D. Conradi, L.J. Garay, E. Recami and P.
Spindel for helpful discussions.


\begin{references}
\bibitem{Remo}  R.Garattini, {\it Vacuum Energy in Ultralocal Metrics for TT
tensors with Gaussian Wave Functionals, }gr-qc/9508060.

\bibitem{Garay}  L.J. Garay, Phys. Rev. D {\bf 48}, 1710 (1993).

\bibitem{MTW}  C.W.Misner, K.S. Thorne and J.A. Wheeler, {\it Gravitation}
(Freeman, San Francisco, 1973) 842; M.S. Morris and K.S. Thorne, Am. J. Phys.%
{\it \ }{\bf 56 }(1988) 395.

\bibitem{Halliwell}  J.J. Halliwell, ''Introductory Lectures on Quantum
Cosmology''. In {\it Jerusalem Winter School for Theoretical Physics:
Quantum Cosmology and Baby Universes Vol. 7}. S.Coleman, J.B. Hartle, T.
Piran and S. Weinberg, eds. World Scientific, 159-243.

\bibitem{Hawking}  S. W. Hawking, in General Relativity: An Einstein
Centenary Survey, ed. S.W. Hawking and W. Israel (Cambridge, 1970).

\bibitem{Carlip}  S. Carlip, C. Teitelboim, Class. Quantum Grav. {\bf 12}
(1995) 1699-1704.

\bibitem{Remo1}  R. Garattini, in preparation.

\bibitem{GPY}  D.J. Gross, M. J. Perry and L. G. Yaffe, Phys. Rev. D {\bf 25}%
, (1982) 330; P. Ginsparg and M. J. Perry, Nucl. Phys. B {\bf 222} (1983)
245.
\end{references}
\end{document}